\documentclass[12pt,preprint]{aastex}

\def\gs{\mathrel{\raise0.35ex\hbox{$\scriptstyle >$}\kern-0.6em
\lower0.40ex\hbox{{$\scriptstyle \sim$}}}}
\def\ls{\mathrel{\raise0.35ex\hbox{$\scriptstyle <$}\kern-0.6em
\lower0.40ex\hbox{{$\scriptstyle \sim$}}}}

\shorttitle{CI emission in IRAS F10214+4724}
\shortauthors{Papadopoulos}

\begin{document}

\title{A search for CI J=2--1 emission in  IRAS F10214+4724}

\author{Padeli \ P.\ Papadopoulos}
\affil{Institut f\"ur Astronomie, ETH Z\"urich, 8093 Switzerland}

\begin{abstract}
Sensitive    new   observations   of    the   fine    structure   line
$^3$P$_2$$\rightarrow  $  $^3$P$_1$  (J=2--1)  of the  neutral  atomic
carbon CI  ($\nu _{rest}\sim 809  $~GHz) in the strongly  lensed Ultra
Luminous Infrared  Galaxy (ULIRG) IRAS  F10214+4724 at z=2.3  with the
mm/sub-mm telescope James Clerk Maxwel (JCMT) are presented.  These do
not confirm the presence of emission from this line at the flux levels
or  angular extent  previously reported  in the  literature.   The new
2$\sigma  $  upper limits  are:  $\rm  S_{CI}\la  7\ Jy\  km\  s^{-1}$
(central position), and  $\rm \langle S_{CI} \rangle \la  8.5\ Jy\ km\
s^{-1}$   (average   over   the   two  $\rm   [\delta   (RA),   \delta
(Dec)]=[0'',\pm  10'']$  positions).   A  CI  emission  assumed  fully
concomitant with  the bulk of  H$_2$ and confined entirely  within the
strongly lensed  object yields an  upper limit of  $\rm M_{CI}(H_2)\la
1.5   \times  10^{10}\  M_{\odot}$,   compatible  with   the  reported
CO-derived  H$_2$  gas  mass,  within  the uncertainties  of  the  two
methods.    A   comparison   with   the  recent   detection   of   the
$^3$P$_1$$\rightarrow  $ $^3$P$_0$  (J=1--0)  line in  this galaxy  by
Weiss  et al.   (2004)  is made  and  the large  discrepancy with  the
previous CI measurements is briefly discussed.

\end{abstract}


\section{Introduction}

The detection of CO J=3--2  emission in the IRAS F10214+4724 galaxy at
z$\sim $2.3  (Brown \&  Vanden Bout 1991;  Solomon, Downes  \& Radford
1992) was the first  detection of a molecular gas  mass tracer at high
redshifts, initiating  a series of largely fruitful  efforts to detect
CO lines  in QSOs (e.g.  Barvainis  et al.  1994; Omont  et al.  1996;
Ohta et al.  1996), sub-mm bright galaxies (e.g.  Frayer et al.  1998;
Neri  et al.  2003),  and Ly-break  galaxies (Baker  et al.   2004) at
increasingly  high redshifts.   The  current record  stands at  z$\sim
$6.42 (Walter  et al.  2003),  dramatically expanding the  domain over
which such standard  ISM probes can be used to  deduce H$_2$ mass, its
average physical  conditions, and (when  sufficient angular resolution
is  available) to  provide information  about total  versus  H$_2$ gas
mass.  On occasion such observations reveal that, unlike local ULIRGs,
high-z starbursts can have star formation activity extending over tens
of kiloparsecs, fueled by large reservoirs of molecular gas whose mass
dominates the total dynamic mass (Papadopoulos et al. 2000).

The importance  of detecting  the two CI  lines at any  redshift stems
from extensive  evidence for fully concomitant $  ^{12}$CO, $ ^{13}$CO
and  CI  emission  in   Galactic  molecular  clouds  (Keene  1997  and
references  therein;  Ikeda  et  al.    2002).   Thus  CI  can  be  an
alternative,  optically  thin,  molecular  gas tracer,  which  remains
sensitive also  for gas with low  excitation conditions (Papadopoulos,
Thi, \& Viti 2004).  Previous  work reported the presence of CO J=3--2
and CI J=1--0,  2--1 emission in the strongly  lensed IRAS F10214+4724
possibly extending  in regions around it and  beyond the amplification
power of the gravitational lens  (Brown \& Vanden Bout 1991, 1992a,b).
This  raised the  possibility of  a  massive amount  of molecular  gas
reservoir  present near this  high-z galaxy,  but follow-up  CO J=3--2
observations  assigned all the  observed CO  emission to  the strongly
lensed object  (Downes, Solomon  \& Radford 1995)  and this  issue was
considered  resolved  (see  Radford  et  al.   1996  for  a  summary).
Observations of the CI J=1--0 line in IRAS F10214+4724 and a few other
high-z objects have been recently reported in the literature where the
emission from  the former is found  to be $\sim 10$  times weaker than
originally reported (Weiss et al.   2004).  Here it must be emphasized
that {\it CI  observations at any redshift are  important on their own
right and are not a trivial variation of CO observations since CI line
emission  remains luminous  also for  low-excitation gas  that  may be
underluminous in  the CO $  J+1\rightarrow J,\ J+1\geq 3$  lines} (see
e.g.  Kaufman  et al.  1999).   This provided the main  motivation for
the follow-up  observations to  search for CI  J=2--1 emission  in and
around this high-z  ULIRG that are presented in  this paper.  A Hubble
constant  of  $\rm  H_{\circ}=75\   Mpc^{-1}\  km\  s^{-1}$  and  $\rm
q_{\circ}=1/2$ are assumed throughout.

\section{Observations and results}

The 15-meter James Clerk Maxwel Telescope (JCMT) on Mauna Kea, Hawaii,
was utilized to observe the CI J=2--1 line ($\rm \nu _{rest}=809.3432$
GHz) towards  IRAS F10214+4724 ($\rm  z=2.2854$ from CO lines;  et al.
1995) in numerous observing sessions, namely in January 17th, 20-22nd,
2003, and during  several sessions between 15th and  28th of June, and
on July 7th  2004.  During the 2003 observing  period the DSB receiver
A3 yielded  typical effective system temperatures  of $\rm T_{sys}\sim
350$~K  and  somewhat higher  ones,  $\rm  T_{sys}\sim (450-600)\  K$,
during the  2004 period.  In the  former period A3 was  tunned at $\rm
\nu   _{cen}(z)=246.4954$~GHz    and   the   Digital   Autocorrelation
Spectrometer (DAS) was set at  its widest bandwidth of 1.84 GHz ($\sim
\rm 2240\  km\ s^{-1}$),  while in the  latter a bandwidth  of 920~MHz
($\rm \sim 1120\ km\  s^{-1}$), centered at $\rm \nu _{cent}=246.3454$
GHz (the exact  line center as expected from the  CO lines), was used.
The wide bandwidth  mode of the DAS was initially used  to allow for a
CI line wider than expected from the high-J CO transitions (since more
gas may be  excited in CI) while maintaining  ample baseline to define
the zero level\footnote{The receiver tunning in the wide band DAS mode
was deliberately  set slightly away  from the expected line  center in
order to avoid positioning the line  in the joining section of the two
spectrometer  sectors that  form  the bandwidth  in  this mode.   This
proved  to be  an unecessary  precaution since  ample  channel overlap
makes for  a smooth  joining of  the two DAS  sectors, evident  in the
co-added spectra in  Figures 1, 2}.  The telescope  pointing and focus
were  checked regularly  using Saturn  and the  spectral-line standard
IRC+10216 (for spectral-line pointing)  and the estimated rms pointing
error was $\sim 3''$ ($\sim $1/7 of the $\rm \theta _{HBPW}=21''$ beam
at this  frequency).  Frequent observations of IRC+10216  at CO J=2--1
verified  the proper  receiver  tunning and  yielded  a spectral  line
calibration uncertainty of $\sim  15\%$.  Rapid beam switching at $\nu
_{sw}\sim  (1-2)$  Hz and  an  azimuthal  throw  of $\sim  60''$  were
employed.  Finally, in  order to search for the  purported extended CI
emission along the North-South direction (Brown \& Vanden Bout 1992b),
the points  $\rm (\Delta  \alpha, \Delta \delta)=(0'',  \pm10'')$ were
also observed  during the 2003  observing period.  The  final co-added
spectra are shown in Figures 1 and 2 respectively.

The quantity  $\rm I=\int  _{\Delta V} \Delta  T^* _A dV$,  where $\rm
 \Delta T^* _A = T^* _A(V)-T^* _{A,bas}$ is the spectral line profile,
 has a thermal rms error of

\begin{equation}
\rm \sigma (I)=\sqrt{N_{\Delta V}}\left[1+\frac{1}{N_{bas}}\right]^{1/2} 
\delta T^* _{A,ch} \Delta V_{chan},
\end{equation}

\noindent
where  $\rm N_{\Delta  V}=\Delta  V/\Delta V_{ch}$  is  the number  of
channels  $\rm \Delta  V_{ch} $  comprising $\rm  \Delta V$,  and $\rm
N_{bas}$ is the total number  of channels used to define the line-free
offset $\rm  T^* _{A,bas}$ ($\rm  N_{bas}/2 $ channels placed  on each
side of the line). The rms error per channel $\rm \delta T^* _{A, ch}$
is assumed uniform over the band.

From the spectra shown in Figures  1 and 2 it is estimated $\rm \sigma
(  I_{(0,0)}) =  0.14\ K\  km\ s^{-1}  $, and  $\rm \delta  (\langle I
\rangle) = 0.17\  K\ km\ s^{-1} $ for the average  spectrum of the two
$(0'',  \pm 10'')$  offset positions.   The FWZI  of the  CI  line was
assumed to be  $\rm \Delta V\sim 2\times \Delta  V_{FWHM}\sim 400\ km\
s^{-1}$ (from multiple, high S/N, CO line observations, Radford et al.
1996).  Upper  limits  can  be  alternatively expressed  in  terms  of
velocity-integrated flux densities from

\begin{equation}
\rm S_{CI}=\int _{\Delta V} S_{\nu } dV = 
\frac{8 k_B}{\eta _a \pi D^2}\int _{\Delta V} \Delta T^* _A dV=
\frac{15.6 (Jy/K)}{\eta _a}\int _{\Delta V} \Delta T^* _A dV,
\end{equation}

\noindent
where  $\eta _a=0.63$  is  the  aperture efficiency  of  JCMT at  this
frequency  (a point  source  was  assumed with  a  source size:  $\leq
1.5''$, Downes  et al.  1995).   The 2$\sigma $ upper  limits obtained
from the  present observations are:  $\rm S_{CI}(0'',0'')\leq 7\  Jy \
km\ s^{-1} $ and $\rm \langle S_{CI}(0'',\pm 10'')\rangle \leq 8.5\ Jy
\ km\ s^{-1}  $ (where a $\sim 15\%$  line calibration uncertainty was
also added in quadrature along with the thermal rms error).

\subsection{A comparison with the J=1-0 transition and an upper limit on M(H$_2$)}

The galaxy  F10214+4724 has  had several of  its CO lines  and thermal
dust continuum frequencies detected  (Solomon Downes, \& Radford 1992;
Downes, Solomon,  \& Radford 1995 and references  therein), from which
$\rm T_k=(50-65)$  K, and $\rm n(H_2)\ga 5\times  10^{3}\ cm^{-3}$ are
deduced. Recently  Weiss et al. (2004)  have measured the  flux of the
J=1--0 line towards the central position using the IRAM 30-m telescope
and found it to be $\rm  S_{CI}(1-0)= (1.6\pm 0.2) Jy\ km\ s^{-1}$. In
principle measurements  of both CI  J=2--1 and J=1--0 lines  can yield
constraints  on the  gas  excitation properties  independent of  those
deduced from the CO  lines.  Indeed, assuming a collisional excitation
of the CI 3-level system (and that the lines are optically thin) it is

\begin{equation}
\rm \frac{S_{CI}(2-1)}{S_{CI}(1-0)}=\frac{A_{21}}{A_{10}}\
\frac{Q_{21}(n, T_k)}{Q_{10}(n, T_k)}=3.38\  \frac{Q_{21}(n, T_k)}{Q_{10}(n, T_k)},
\end{equation}

\noindent
  where   $\rm    A_{21}=2.68\times   10^{-7}\   s^{-1}$    and   $\rm
A_{10}=7.93\times  10^{-8}\ s^{-1}$ are  the Einstein  coefficients of
the two transitions, and $\rm Q_{ul}(n, T_k)= N_u/N_{tot}$ express the
relative  population levels (their  full expressions  can be  found in
Papadopoulos  et al. 2004).   The upper  limit on  the CI  J=2--1 line
emission reported here and the  measurement by Weiss at al. for J=1--0
yield $\rm Q_{21}/Q_{10}\la 1.29 $, which is fully compatible with the
values expected for  the conditions reported for the  molecular gas in
this  galaxy  (Downes  et   al.   1995)  but  hardly  constraining  by
itself. However  somewhat more sensitive CI  J=2--1 measurements would
yield  non-trivial constraints, e.g.   for $\rm  Q_{21}/Q_{10}\sim 1/2
\times  1.29=0.64 $  (corresponding to  $\sim 1\sigma  $ limit  of the
current measurements) one finds $\rm  T_k\sim 40\ K$ (for LTE) to $\rm
T_k\sim 55\ K $ (for $\rm n=5\times 10^3\ cm ^{-3}$), which now probes
the temperature range deduced from CO and dust measurements.

 Assuming  that CO,  dust and  CI  emission are fully concomitant  and
tracing  the  same H$_2$  gas  reservoir  allows  $\rm S_{CI}$  to  be
expressed in terms of H$_2$ mass by using

\begin{equation}
\rm \frac{M_{ci}(H_2)}{M_{\odot}}=8.75\times 10^{10}
\frac{(1+z-\sqrt{1+z})^2}{1+z}
\left[\frac{X_{CI}}{10^{-5}}\right]^{-1}
\rm \left[\frac{A_{21}}{10^{-7} s^{-1}}\right]^{-1} Q_{21} ^{-1}
\left[ \frac{S_{CI}(2-1)}{Jy\ km\ s^{-1}}\right],
\end{equation}

\noindent
(see  e.g. Papadopoulos et  al.  2004)  where $\rm  Q_{21}(n, T_k)\sim
0.33$ (for $\rm T_k = 65\ K$ and $\rm n=5\times 10^3\ cm^{-3}$; Downes
et  al.  1995).  An  abundance  of  $\rm X_{CI}=[C]/[H_2]\sim  3\times
10^{-5}$ is adopted, identical to the one used by Weiss et al.  (2003)
in their  H$_2$ gas mass estimate  in the Cloverleaf  QSO.  However an
uncertainty of a factor of two  in $\rm X_{CI}$ is to be expected over
the  range of  ISM  conditions  (higher values  more  likely for  star
forming environments).

 For a lensing factor $m_{CI}=m_{CO}\sim 10$ (Downes et al.  1995) the
 reported  upper  limit  on  $\rm  S_{CI}(0'',0'')$  translates  to  a
 molecular  gas  mass upper  limit  of  $\rm M_{ci}(H_2)\la  1.5\times
 10^{10}\  M_{\odot}$. The  latter is  compatible with  the CO-deduced
 molecular  gas  mass  of  $\rm M(H_2)=2\times  10^{10}\  M_{\odot  }$
 (Downes  et al.   1995), within  the expected  uncertainties  of both
 methods. It must  be mentioned that the CI  J=2--1 emission, with its
 less  demanding  excitation   requirements,  can  be  more  spatially
 extended  than that  of the  CO J+1$\rightarrow  $J, $\rm  J+1\geq 3$
 lines (see e.g.   Kaufman et al.  1999, their Figures  10 and 13) and
 then any  differential lensing  would act to  raise the value  of the
 reported  $\rm M_{ci}(H_2)$  upper limit  (less amplification  of the
 more extended emission).

\subsubsection{The discrepancy with past observations}

The deduced upper  limits are much lower than  the $\rm S_{CI}$ values
reported  by Brown \&  Vanden Bout  (1992b) which  used the  NRAO 12-m
Telescope,  with  a   beam  of  $\sim  28''$  (HPBW),   to  find  $\rm
S_{CI}(0'',0)=150$ Jy  km s$  ^{-1}$. There is  no indication  of such
strong  emission in  the data  presented here,  which would  have been
detected at a S/N$\sim 40$. The possibility that most of such emission
is distributed in the area between the $21''$ and $28''$ beam areas of
the  JCMT  and the  12-m  telescope  is  remote.  Indeed  the  average
spectrum of the two offset positions $(0'', \pm 10'')$ (Fig.  2) where
the purported extended CI emission was claimed to be the strongest, do
not  confirm its  presence  at the  levels  previously reported.   The
latter disagreement  can be wholy  attributed to the poor  quality and
low  S/N ratio  of the  associated spectra  taken with  the  IRAM 30-m
telescope.  However in their 12-m  data a line towards the $(0'',0'')$
position appears detected at a  sufficient S/N ratio.  For this set of
their  measurements  baseline/spectrometer  instabilities are  a  more
likely cause for the  discrepancy\footnote{In that respect it is worth
noting  that   their  NRAO  12-m  observations   were  obtained  under
relatively  humid conditions  ($\rm  T_{sys}\sim 1250$  K), using  two
different receiver  tunnings in  order to cover  the line, and  with a
chop throw that is four times  larger than the one employed here.  All
these  are expected  to make  their measurements  especially  prone to
weather/instrumentally-induced   baseline  instabilities.}.    In  the
present dataset an intermittent baseline instability affected the data
taken over a long observing session during the 2003 period and appears
as a  spectral line detected at  a S/N=4-5 (Figure  3).  This ``line''
appears  strongly only  in  a handful  of  the spectra  making up  the
co-added spectrum,  and its  low-level intermittent nature  along with
the  otherwise  normal  thermal  noise  superimposed on  it  made  the
affected  spectra   difficult  to  identify  in   the  data  reduction
processes.   This was made  possible only  after coarse  averaging and
inspection of all individual spectra  and the acquisition of more data
spread over several observing sessions and with different spectrometer
setups.

\section{Conclusions}

 The results of this work can be summarized as follows:

\noindent
 1. New sensitive observations of the  CI J=2--1 line emission in IRAS
F10214+4724  at $\rm z\sim  2.3$ do  not confirm  its presence  at the
intensity or the angular extent previously reported in the literature.

\noindent
2. Under the assumption that CI  emission traces the bulk of the H$_2$
gas mass located in the  strongly-lensed object an upper limit of $\rm
M_{CI}(H_2)\leq  1.5\times  10^{10}\ M_{\odot}$  is  deduced. This  is
compatible with the CO-derived H$_2$ mass, within the uncertainties of
the two methods.  However any differential lensing effects between the
easier-to-excite and possibly more  extended CI emission (with respect
to CO J=3--2) would act to raise the CI-derived H$_2$ mass limit.

\noindent
3. A comparison with the recently reported CI J=1--0 line intensity in
   this object by Weiss et al.  (2004) allows an upper limit on the CI
   (2--1)/(1--0) line ratio to be placed.  Its value, while compatible
   with the average CO-derived  excitation conditions of the molecular
   gas,  does   not  offer  any  useful   independent  constraints  by
   itself.  Nevertheless only a  modest improvement  of any  future CI
   J=2--1 measurements  (a factor  of 2) is  needed in order  to yield
   a good such independent constraint.

\noindent
4. The  discrepancy  between the  hereby  presented measurements  with
those  of Brown  \&  Vanden Bout  is  attributed to  the  low S/N  and
baseline  instabilities affecting  their data.   The latter  source of
error  can be particularly  insidious as  demonstrated by  an affected
part of dataset presented here  where such an instability could easily
be construed as a line detected at $\sim (4-5)\sigma$.

\acknowledgments

I thank the JCMT Operator  Jim Hoge whose determination made this work
possible, often  under difficult conditions.  Axel  Weiss is gratefuly
acknowledged for  his mistrust of  the spectral ``line''  appearing on
the first  dataset.  This has  been proven crucial in  re-reducing the
original dataset  much more carefuly  as well as acquiring  more data,
both of which  led to the revision of the original  claim of CI J=2--1
detection in F10214+4724 and the hereby reported upper limits.

\clearpage

\begin{figure}
\includegraphics[angle=180,scale=1.00]{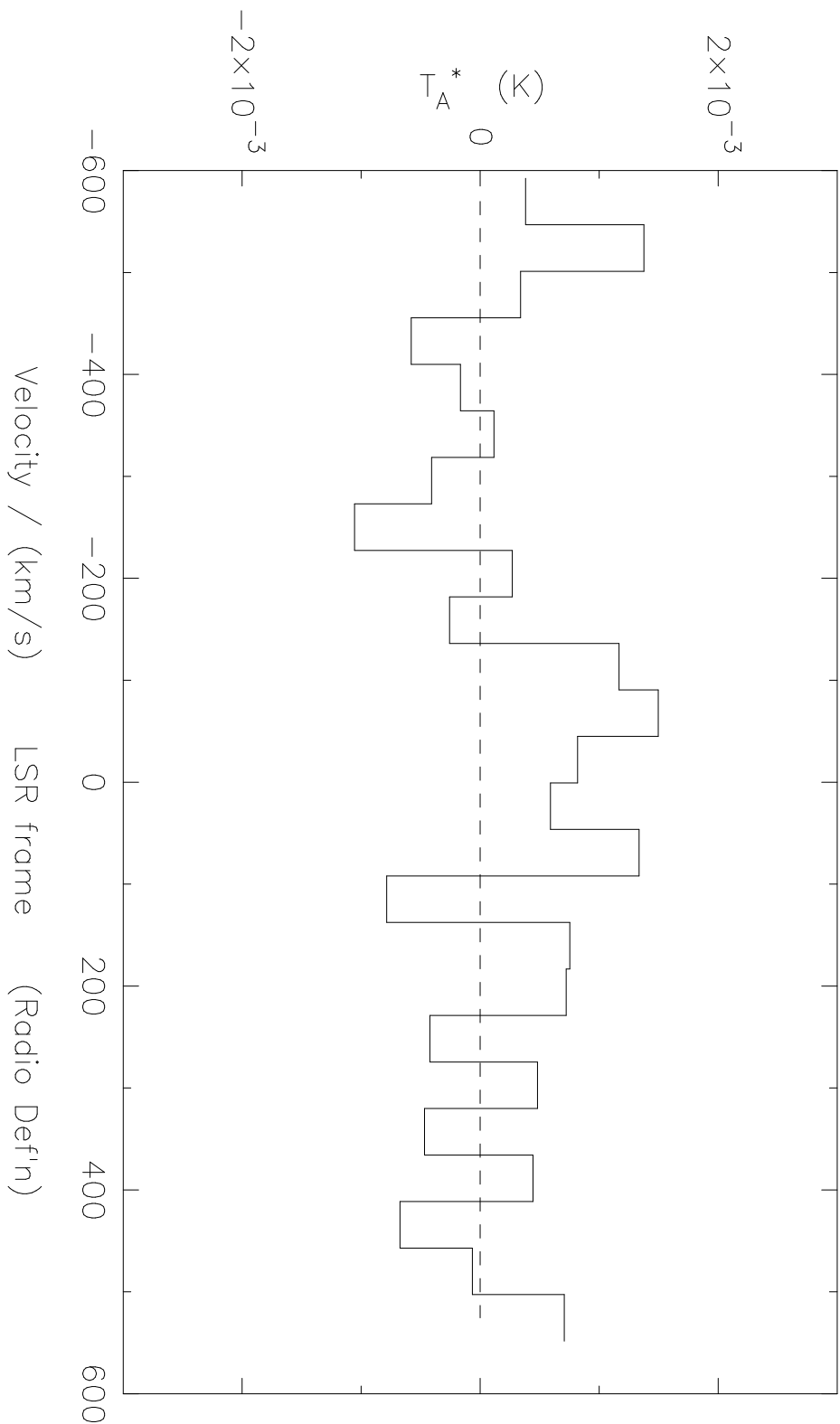}
\caption{The CI  J=2--1 spectrum on  the central position  $\rm \alpha
_{1950}:  10 ^h\  21  ^{m}\ 31.1  ^s$,  and $\rm  \delta _{1950}:  +47
^{\circ}\ 24  ^{'}\ 23.0 ^{''}$  of IRAS F10214+4724 (Brown  \& Vanden
Bout 1992).  The estimated channel-to-channel rms noise is $\rm \delta
T^*  _{A,  ch} =  1\  mK$,  at a  velocity  resolution  of $\rm  \Delta
V_{chan}=46\ km\  s^{-1}$. The line-free  offset $\rm T^* _{A,  bas} $
was estimated  from $\rm N_{bas}/2= 7$  channels, placed symmetrically
around the  central 400  km s$ ^{-1}$  of the bandwidth.  The velocity
offsets are  estimated from a  central frequency of 246.3454  GHz, the
expected line center (see text). }
\end{figure}

\begin{figure}
\includegraphics[angle=180,scale=1.00]{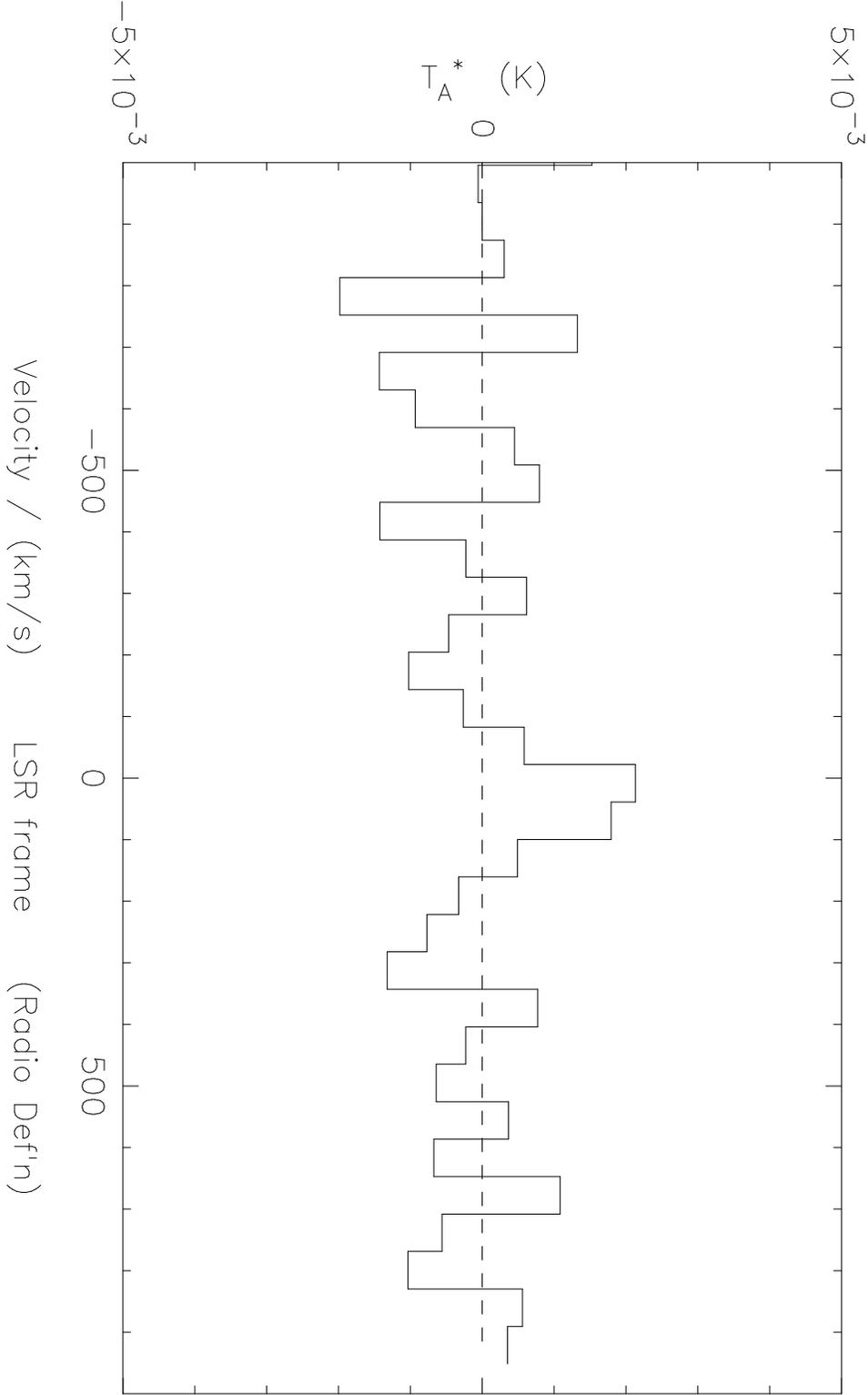}
\caption{The  average CI  J=2--1  spectrum of  the positions  $(\Delta
\alpha,   \Delta   \delta)$:  $(0'',   \pm   10'')$.   The   estimated
channel-to-channel rms  noise is  $\rm \delta  T^* _A =  1\ mK$,  at a
velocity  resolution of  $\rm  \Delta V_{chan}=61\  km\ s^{-1}$.   The
line-free  offset  $\rm  T^*  _{A,  bas} $  was  estimated  from  $\rm
N_{bas}/2= 8$ channels, placed symmetrically around the central 400 km
s$ ^{-1}$ of the bandwidth.  The velocity offsets are estimated from a
central  frequency of  246.3454  GHz, the  expected  line center  (see
text). }
\end{figure}

\begin{figure}
\includegraphics[angle=180,scale=1.00]{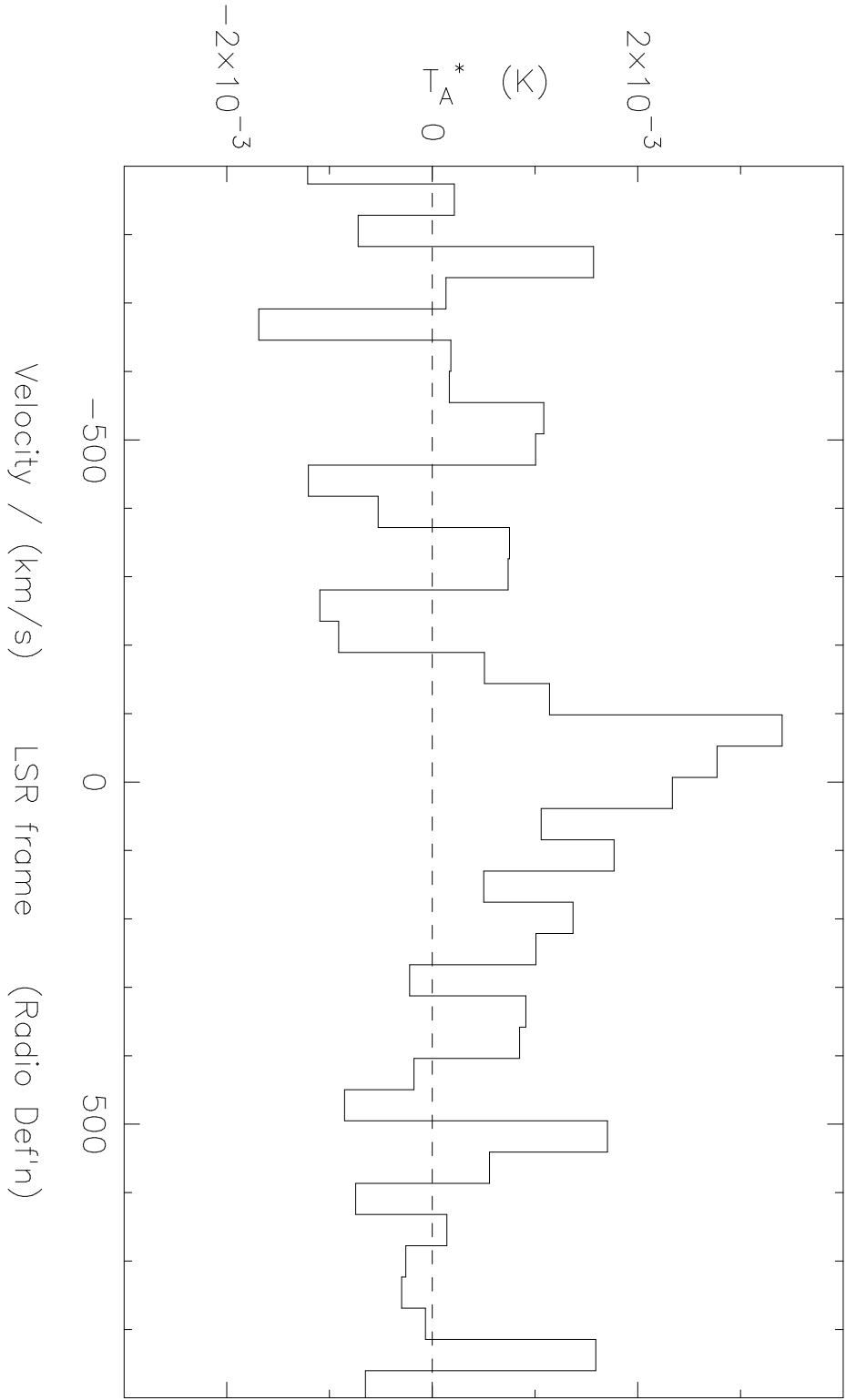}
\caption{A 9-hour long spectrum,  averaged over the positions $(\Delta
\alpha,  \Delta  \delta)$: $(0'',  0'')$,  $(0'',\pm  10'')$, made  by
including data containing a  low-level baseline offset (see text), the
resulting spurious line has a S/N=4-5.}
\end{figure}

\end{document}